\documentclass[aps,prc,floatfix,nofootinbib,preprintnumbers,superscriptaddress,longbibliography,notitlepage,twocolumn]{revtex4-2}
\usepackage[dvipsnames,svgnames,x11names]{xcolor}
\usepackage{epsfig}
\usepackage{graphicx}
\usepackage{stmaryrd}
\usepackage{amssymb,tabularx,dcolumn}
\usepackage{supertabular,ltxtable}
\usepackage{longtable}
\usepackage{amsmath}
\usepackage{amstext}
\usepackage{float}
\usepackage{color}
\usepackage{bm}
\usepackage{hyperref}
\hypersetup{breaklinks=true,colorlinks=true,linkcolor=blue,citecolor=blue,filecolor=magenta,urlcolor=cyan}
\usepackage{mathtools}
\usepackage{multirow}
\usepackage{makecell}
\usepackage[utf8]{inputenc}
\usepackage{tabu}
\usepackage{braket}
\usepackage{subfigure}
\usepackage{enumitem}
\usepackage[english]{babel}
\definecolor{pastelgray}{rgb}{0.81, 0.81, 0.77}
\definecolor{beaublue}{rgb}{0.9, 0.9, 0.93}

\newcommand{\midrule}{\hline}
\newcommand{\bottomrule}{\hline}

\begin{document}
\title{Uncertainty Quantification of the $^{76}$Ge Neutrinoless Double-Beta Decay Nuclear Matrix Element}
\author{M. Horoi}
\affiliation{Department of Physics, Central Michigan University, Mount Pleasant, MI 48859, USA}
\affiliation{International Center for Advanced Training and Research in Physics (CIFRA), Magurele, Romania}
\author{A. Neacsu}
\affiliation{International Center for Advanced Training and Research in Physics (CIFRA), Magurele, Romania}
\affiliation{Institute of Atomic Physics (IFA), Magurele, Romania}

\begin{abstract}
The experimental pursuit of neutrinoless double-beta decay ($0\nu\beta\beta$) constitutes one of the most compelling avenues for probing lepton-number violation and exploring physics beyond the Standard Model. Within this landscape, $^{76}$Ge has consistently ranked among the most promising isotopes for current and next-generation bolometric and liquid-scintillator experiments, notably GERDA and LEGEND. In the present work, we adapt a rigorous statistical protocol previously established for $^{48}$Ca~\cite{Horoi-prc22} and $^{136}$Xe~\cite{Horoi-Xe-2023} to the $^{76}$Ge system, utilizing a valence configuration that aligns with our recent investigation of $^{82}$Se~\cite{Neacsu-Symmetry-2024}. Our methodology introduces systematic, bounded fluctuations to the two-body matrix elements of established effective interactions, subsequently monitoring how these perturbations propagate through a suite of low-energy nuclear observables. Special emphasis is placed on the $0\nu\beta\beta$ nuclear matrix element (NME), whose theoretical uncertainty currently dominates the interpretation of experimental half-life limits. By integrating these simulated variations into a Bayesian Model Averaging framework and benchmarking against empirical spectroscopic data, we derive a constrained probability distribution for the NME. The resulting analysis yields a central value of 2.46 with an associated standard deviation of 0.25, thereby quantifying the intrinsic theoretical spread within the interacting shell model approach. Furthermore, we perform a comprehensive correlation analysis across all computed observables to evaluate internal consistency, identify non-trivial structural dependencies, and establish benchmarks that may guide the refinement of future effective interactions.
\end{abstract}

\maketitle

\section{Introduction}

The Standard Model of particle physics rests upon several foundational conservation laws, among which lepton number and lepton flavor play central roles. The discovery of neutrino oscillations~\cite{SuperKamiokande1998,SNO2001}, a phenomenon theoretically anticipated in the late 1950s~\cite{Pontecorvo1958, Pontecorvo1968}, definitively demonstrated that lepton flavor is not an exact symmetry of nature. This paradigm-shifting result, recognized with the 2015 Nobel Prize in Physics~\cite{Nobel2015}, highlights the incomplete status of the Standard Model and motivates the search for new dynamical mechanisms. In ordinary beta decay ($\beta^{-/+}$), lepton number conservation is maintained through the simultaneous emission of a charged lepton and its corresponding (anti)neutrino. Although particle-antiparticle symmetry has been verified to extraordinary precision in controlled laboratory environments, the observable cosmos exhibits a profound matter-antimatter asymmetry. Reconciling this discrepancy requires theoretical extensions that permit fundamental symmetry violations, often manifesting through rare nuclear transitions.

The hypothetical neutrinoless double-beta decay ($0\nu\beta\beta$) occupies a unique position at the intersection of nuclear physics, neutrino phenomenology, and beyond-Standard-Model theory. While the standard two-neutrino mode ($2\nu\beta\beta$) has been experimentally confirmed in eleven isotopes~\cite{Barabash2020}, the $0\nu\beta\beta$ channel remains unobserved. This decay process involves the simultaneous transformation of two neutrons into two protons with the emission of only two electrons, thereby violating lepton number conservation by $\Delta L = 2$. Detection of this mode would provide unambiguous evidence for the Majorana character of neutrinos, wherein the neutrino serves as its own antiparticle~\cite{SchechterValle1982,Hirsch2006}. Such a discovery would also constrain the absolute neutrino mass scale, inform the structure of the lepton sector, and offer indirect access to high-energy physics scales that remain beyond the reach of terrestrial colliders~\cite{Avignone2008,Vergados2012}.

Theoretical frameworks incorporating Type I or Type II seesaw mechanisms naturally predict Majorana neutrino masses and, consequently, $0\nu\beta\beta$ transitions. Moreover, the observation of this decay could reveal the presence of right-handed weak currents, a hallmark of left-right symmetric extensions~\cite{Rodejohann2012}. Beyond neutrino physics, $0\nu\beta\beta$ decay serves as a low-energy probe for Grand Unified Theories (GUTs) and dark matter models, which often predict new degrees of freedom at energy scales inaccessible to direct production. Experimental collaborations worldwide are continuously advancing detector technologies, pushing half-life sensitivity beyond $10^{26}$ years~\cite{GERDA2020,Majorana2023,EXO200-2019,PhysRevLett.130.051801}. The most restrictive bounds to date, approaching $2 \cdot 10^{26}$ years, have been established using $^{76}$Ge~\cite{GERDA2020} and $^{136}$Xe~\cite{PhysRevLett.130.051801}.

Within the light Majorana neutrino exchange mechanism, the inverse half-life is factorized into three distinct components: a precisely calculable leptonic phase-space factor (PSF), a lepton-number-violating parameter encoding the effective Majorana mass, and the nuclear matrix element (NME) that encapsulates the complex many-body dynamics of the parent and daughter nuclei~\cite{Doi1983,Doi1985,SuhonenCivitarese1998}. While PSF values are known to high accuracy~\cite{Nitescu:2024ppf, Nitescu:2024rsh, Nitescu:2023sry, Nitescu:2021pdq, Nabi:2019usr,MireaPahomi2015, StoicaMirea2013, Kotila2012} and the LNV parameter will be extracted upon experimental discovery, the NME currently represents the dominant source of theoretical uncertainty. A variety of nuclear structure methodologies have been deployed to evaluate these matrix elements, including the interacting shell model (ISM)~\cite{Caurier1990,Caurier1996, Caurier2005, HoroiStoicaBrown2007, HoroiStoica2010, Horoi2013, HoroiBrown2013, SenkovHoroi2014, NeacsuHoroi2015, NeacsuHoroi2016,18ho035502}, the proton-neutron quasiparticle random phase approximation (pn-QRPA)~\cite{SuhonenCivitarese1998, Simkovic1999, Stoica2001, Rodin2006, KortelainenSuhonnen2007, Faessler2012, SimkovicRodin2013}, the interacting boson approximation (IBA)~\cite{Barea2009, Barea2013}, energy density functionals~\cite{Rodriguez2010}, projected Hartree-Fock-Bogoliubov approaches~\cite{Rath2013}, coupled-cluster techniques~\cite{Novario2021}, in-medium generator coordinate methods~\cite{Yao2020}, and valence-space in-medium similarity renormalization group calculations~\cite{Belley2021}. Each framework entails distinct computational trade-offs, correlation treatments, and model-space limitations, which collectively contribute to the broad spread of published NME values~\cite{EngelMenendez2017}. Comprehensive reviews have recently addressed these discrepancies and evaluated the Bayesian discovery potential for upcoming experiments~\cite{Rodejohann2019review}.

Among the available approaches, the interacting shell model stands out for its systematic treatment of configuration mixing and its minimal reliance on phenomenological tuning. Predictions generated by independent groups employing different effective interactions typically converge, and the ISM successfully anticipated $2\nu\beta\beta$ half-lives prior to their experimental verification~\cite{Retamosa1995,Balysh1996}. These attributes make the shell model particularly well-suited for quantifying theoretical uncertainties in $0\nu\beta\beta$ transitions.

The remainder of this manuscript is structured as follows: Section~\ref{model} details the statistical protocol and computational setup. Section~\ref{results} presents the Bayesian-averaged NME distribution for $^{76}$Ge, accompanied by a thorough correlation analysis. Section~\ref{conclusions} summarizes our findings and outlines prospective theoretical developments. While earlier investigations validated effective interactions within the $jj55$ valence space~\cite{NeacsuHoroi2015,NeacsuHoroi2016}, the present work extends this statistical methodology to the $jj44$ configuration, with particular attention paid to observable interdependencies and their experimental benchmarks.

\section{Statistical Framework and Methodology} \label{model}

Building upon the statistical protocols established for $^{48}$Ca in the {\it fp} space~\cite{Horoi-prc22} and $^{136}$Xe in the {\it jj55} space~\cite{Horoi-Xe-2023}, we now apply the same ensemble-based approach to the {\it jj44} model space (also labeled $f_{5}pg_{9}$), which is directly applicable to the $^{76}$Ge $\rightarrow$ $^{76}$Se transition as well as the $^{82}$Se system. This valence configuration is constructed upon a $^{56}$Ni inert core, with active nucleons occupying the $1p_{3/2}$, $1p_{1/2}$, $0f_{5/2}$, and $0g_{9/2}$ orbitals. 

The computational recipe follows the methodology outlined in Ref.~\cite{Horoi-Xe-2023}. We begin with three well-established effective Hamiltonians and generate perturbed ensembles by uniformly varying their two-body matrix elements (TBME) within a $\pm10\%$ window around the original values. This perturbation amplitude is deliberately chosen to remain within empirically motivated uncertainty bounds while preventing unphysical overfitting~\cite{Brown2006}. Single-particle energies (SPE) are held fixed throughout the sampling procedure. Given the steep computational scaling associated with full diagonalizations in the $jj44$ space, we restrict our statistical sampling to 200 perturbed Hamiltonians per starting interaction. No model-space truncation is applied, ensuring that all relevant configuration mixing is captured.

The three baseline interactions employed in this study are JUN45~\cite{JUN45}, GCN2850~\cite{MenendezNPA818}, and JJ44b~\cite{JJ44b-int}. These effective forces originate from the Bonn-C nucleon-nucleon potential and are subsequently refined through fits to experimental spectroscopic data. Detailed construction procedures, including treatments of missing orbitals and Ikeda sum-rule satisfaction, are documented in Ref.~\cite{JUN45}. In the $jj44$ environment, the $^{56}$Ni core exhibits significant softness, meaning that low-lying states in neighboring nuclei cannot be interpreted as pure single-particle excitations. Consequently, SPE values must be empirically adjusted during Hamiltonian parametrization rather than extracted directly from one-nucleon transfer data.

For each perturbed Hamiltonian, we compute a comprehensive set of low-energy observables that feed into the statistical analysis. Beyond the $0\nu\beta\beta$ NME, these include the $2\nu\beta\beta$ NME, excitation energies of the first $2^+$, $4^+$, and $6^+$ states in both $^{76}$Ge and $^{76}$Se, B(E2)$\uparrow$ transition strengths to the first $2^+$ levels, and Gamow-Teller transition probabilities connecting ground states in $^{76}$Ge/$^{76}$Se to the first $1^+$ state in the intermediate nucleus $^{76}$As. Unlike earlier studies that also incorporated proton occupancies and neutron vacancies~\cite{Horoi-Xe-2023}, we omit these quantities here due to the absence of reliable experimental benchmarks for $^{76}$Ge and $^{76}$Se. The selected observables provide a robust cross-check of Hamiltonian fidelity and enable systematic correlation studies that may guide future interaction refinements.

Calculation details for these quantities are provided in Refs.~\cite{Horoi-prc22,Horoi-Xe-2023,Horoi-universe-2024}. For Gamow-Teller transitions and the $2\nu\beta\beta$ NME, we apply a phenomenological quenching factor $q=0.65$, which is standard in the {\it fp} space for $^{48}$Ca and closely matches values used in the {\it jj55} space for Sn-Te-Xe isotopes with the SVD interaction~\cite{Chong2012}. The sensitivity of results to $q$ variations has been thoroughly examined in Ref.~\cite{Horoi-universe-2024}. In contrast, the $0\nu\beta\beta$ NME ($M_{0\nu}$) is evaluated without quenching, employing the closure approximation to replace intermediate-state energies with an average value. Short-range nucleon correlations are modeled using the Jastrow form parametrized by Miller and Spencer~\cite{Miller1976}. 

We deliberately avoid fine-tuning model parameters to reproduce specific experimental points, as such overfitting compromises the predictive universality of effective interactions and may introduce unphysical distortions in unobserved observables, such as nuclear deformation or shell evolution. Both $^{76}$Ge and $^{76}$Se exhibit pronounced quadrupole deformation. For B(E2) calculations, we adopt the effective charges $e_p=1.5$ and $e_n=1.1$, as recommended for the $jj44$ valence space in Ref.~\cite{JUN45}. These values ensure consistent treatment across all three Hamiltonians and their perturbed variants.

\section{Results and Discussions} \label{results}

To establish a reliable baseline for our statistical ensemble, we first evaluate all observables using the unperturbed starting Hamiltonians and compare them against available experimental measurements. This calibration step validates the computational pipeline before proceeding to the full Bayesian analysis incorporating 200 perturbed Hamiltonians per interaction.

Table~\ref{allstats} summarizes the experimental benchmarks, initial Hamiltonian predictions, and ensemble statistics. The prefixes ``{\bf P}'' and ``{\bf D}'' refer to the parent $^{76}$Ge and daughter $^{76}$Se nuclei, respectively. The columns report the $0\nu\beta\beta$ NME ($M_{0\nu}$), $2\nu\beta\beta$ NME ($M_{2\nu}$), Gamow-Teller strengths (P$GT$, D$GT$), B(E2)$\uparrow$ transition probabilities, and excitation energies for the $2^+$, $4^+$, and $6^+$ states. Experimental uncertainties are listed in the third column; for energy levels, a conservative error of 150 keV is adopted~\cite{Horoi-prc22, Horoi-Xe-2023,Horoi-universe-2024}. Columns 4 through 12 present results from the three baseline Hamiltonians ($gcn_s$, $jun_s$, $jj4_s$), followed by the ensemble means ($\mu_{gcn}$, $\mu_{jun}$, $\mu_{jj4}$) and standard deviations ($\sigma_{gcn}$, $\sigma_{jun}$, $\sigma_{jj4}$).

\begin{widetext}
\begin{center}
\begin{table}

\begin{tabular}{lrrrrrrrrrrr}
\toprule
Observable & Data & Error & gcn & jun & jj4 & $\mu_{gcn}$ & $\sigma_{gcn}$ & $\mu_{jun}$ & $\sigma_{jun}$ & $\mu_{jj4}$ & $\sigma_{jj4}$ \\
\midrule
$M_{0\nu}$ & NA & NA & 2.389 & 2.701 & 2.534 & 2.398 & 0.219 & 2.685 & 0.220 & 2.499 & 0.152 \\
$M_{2\nu}$ & 0.129 & 0.001 & 0.121 & 0.120 & 0.115 & 0.121 & 0.014 & 0.119 & 0.014 & 0.113 & 0.009 \\
P$GT$ & 0.120 & 0.013 & 0.160 & 0.079 & 0.093 & 0.145 & 0.068 & 0.093 & 0.053 & 0.109 & 0.026 \\
P$BE2$ & 0.273 & 0.003 & 0.242 & 0.250 & 0.280 & 0.241 & 0.008 & 0.247 & 0.010 & 0.278 & 0.006 \\
P$E_{2^+}$ & 0.563 & 0.150 & 0.718 & 0.745 & 0.718 & 0.719 & 0.049 & 0.750 & 0.053 & 0.726 & 0.045 \\
P$E_{4^+}$ & 1.410 & 0.150 & 1.585 & 1.637 & 1.653 & 1.583 & 0.089 & 1.640 & 0.091 & 1.667 & 0.084 \\
P$E_{6^+}$ & 2.453 & 0.150 & 2.527 & 2.646 & 2.683 & 2.517 & 0.125 & 2.643 & 0.137 & 2.701 & 0.115 \\
D$GT$ & 0.012 & 0.005 & 0.029 & 0.019 & 0.031 & 0.030 & 0.018 & 0.023 & 0.014 & 0.037 & 0.013 \\
D$BE2$ & 0.432 & 0.010 & 0.321 & 0.315 & 0.369 & 0.320 & 0.016 & 0.314 & 0.024 & 0.363 & 0.016 \\
D$E_{2^+}$ & 0.559 & 0.150 & 0.511 & 0.574 & 0.669 & 0.520 & 0.056 & 0.583 & 0.072 & 0.660 & 0.045 \\
D$E_{4^+}$ & 1.331 & 0.150 & 1.274 & 1.353 & 1.638 & 1.283 & 0.096 & 1.366 & 0.118 & 1.624 & 0.096 \\
D$E_{6^+}$ & 2.262 & 0.150 & 2.173 & 2.286 & 2.780 & 2.180 & 0.124 & 2.299 & 0.147 & 2.765 & 0.147 \\
\bottomrule

\end{tabular}
\caption{The experimental data, the calculated values with the starting Hamiltonians, and the statistics for the relevant 12 observables.} \label{allstats}

\end{table}
\end{center}
\end{widetext}

The ensemble averages closely track the unperturbed Hamiltonian predictions, and the standard deviations remain modest across all observables. This behavior indicates the absence of pathological sensitivity to specific TBME fluctuations and confirms that no hidden phase-transition-like dynamics or dominant parameter subsets distort the results. Such stability reinforces confidence in the robustness of shell-model predictions and their capacity to yield reliable uncertainty estimates. It is worth emphasizing that, despite the numerical agreement among starting Hamiltonians, their underlying parameters differ substantially: SPE variations reach MeV scales for identical orbitals, and both diagonal and off-diagonal TBME exhibit comparable discrepancies.

\begin{center}
\begin{figure*}[htbp]
\includegraphics[clip, trim={2.5cm 3.5cm 6.5cm 4.0cm},width=17cm]{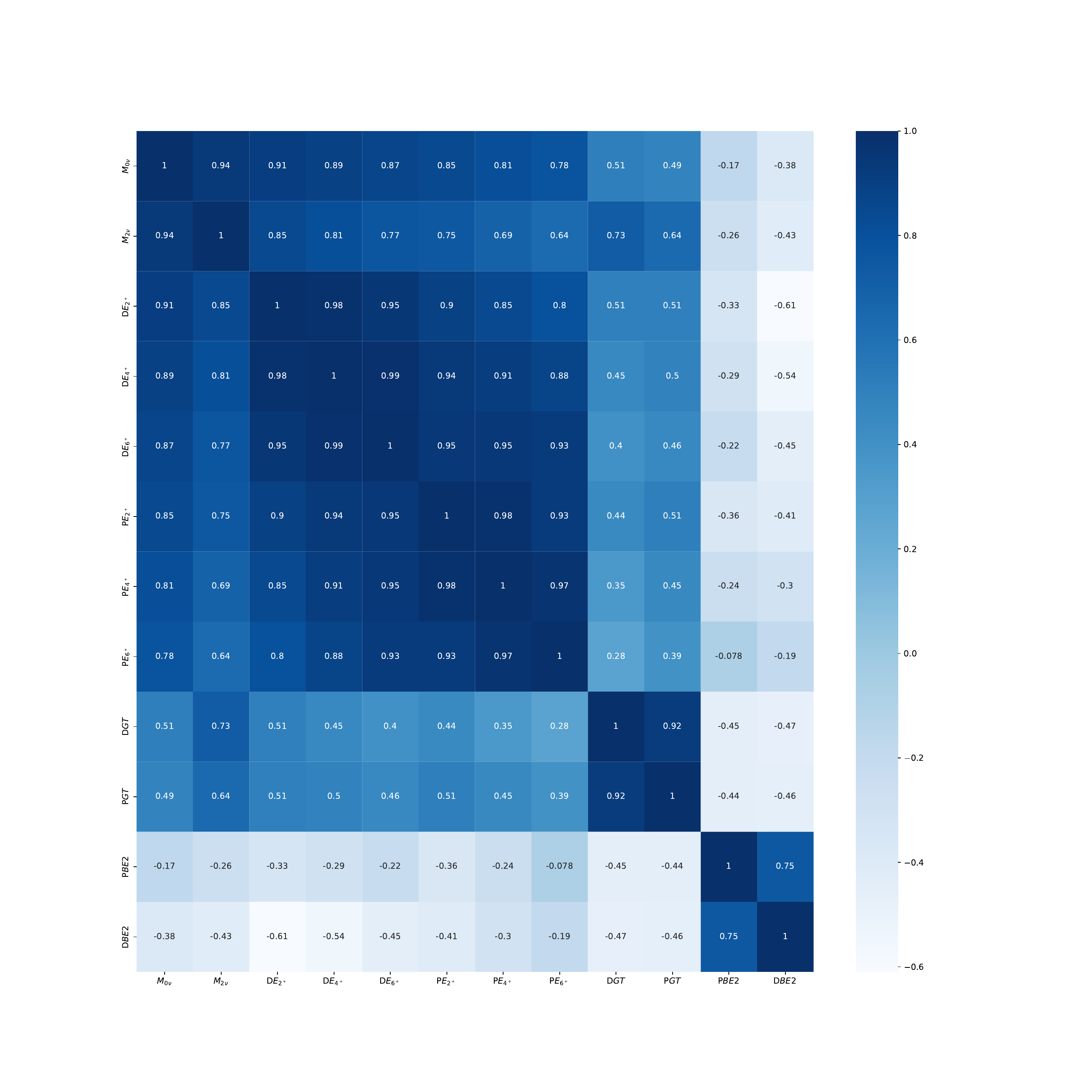}
\caption{The heatmap for all 12 observables when using the GCN2850 effective Hamiltonian. Very similar representations are obtained with the JUN45 and JJ44b Hamiltonians.}
\label{gcn-hm}
\end{figure*}
\end{center}

Figure~\ref{gcn-hm} presents a Pearson correlation heatmap constructed from the ensemble variations. The color scale spans from $+1$ (perfect positive correlation) to $-1$ (perfect anti-correlation), with zero indicating statistical independence. We adopt a threshold of $|\rho| > 0.5$ to flag physically meaningful dependencies. Consistent with earlier studies on $^{48}$Ca and $^{136}$Xe, the $M_{0\nu}$ and $M_{2\nu}$ observables exhibit a correlation coefficient exceeding 0.8, underscoring a robust structural link despite their distinct operator structures: $M_{2\nu}$ is governed by $1^+$ intermediate states, whereas $M_{0\nu}$ integrates contributions from all neutrino-momentum-allowed virtual excitations. Additional strong correlations emerge among excited-state energies and between Gamow-Teller strengths and B(E2) transition rates. Notably, parent-daughter energy level correlations are pronounced, providing confidence for future calculations targeting transitions to excited final states.

\begin{center}
\begin{figure*}[htbp]
\includegraphics[clip, trim={0.2cm 0.2cm 0.0cm 0.0cm},width=20cm]{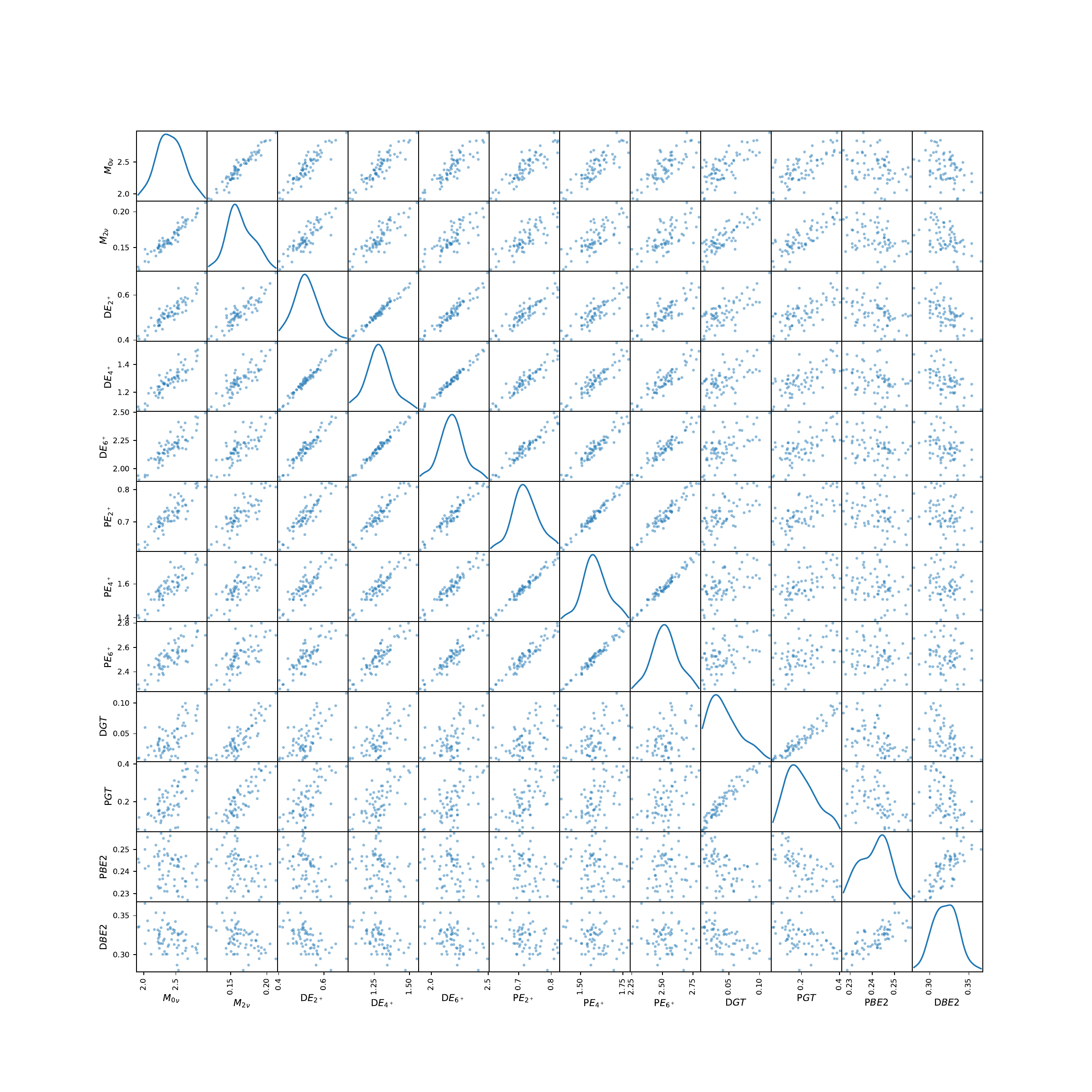}
\caption{Correlation matrix for observables that have correlation factor greater than 0.5, when using the GCN2850 Hamiltonian.}
\label{gcn-corr}
\end{figure*}
\end{center}

For observable pairs satisfying $|\rho| > 0.5$, Figure~\ref{gcn-corr} displays a detailed correlation matrix featuring contour density plots and downsampled bubble scatter representations. The diagonal panels contain histograms of individual observable distributions, which consistently follow Gaussian profiles. This behavior is expected for ensembles generated by bounded random perturbations and confirms the absence of extreme parameter sensitivity or non-linear response regimes.

Visual inspection of Figure~\ref{gcn-corr} reveals that stronger correlations manifest as tighter clustering along the main diagonal, while anti-correlations align with the off-diagonal. Clear examples include the $2^+$ and $4^+$ excitation energies in parent and daughter systems, as well as the $M_{0\nu}$-$M_{2\nu}$ linkage. Conversely, parent B(E2)$\uparrow$ values exhibit anti-correlation with several NME and energy observables, reflecting underlying structural trade-offs in configuration mixing.

\begin{center}
\begin{figure*}[htbp]
\includegraphics[ width=8.5cm]{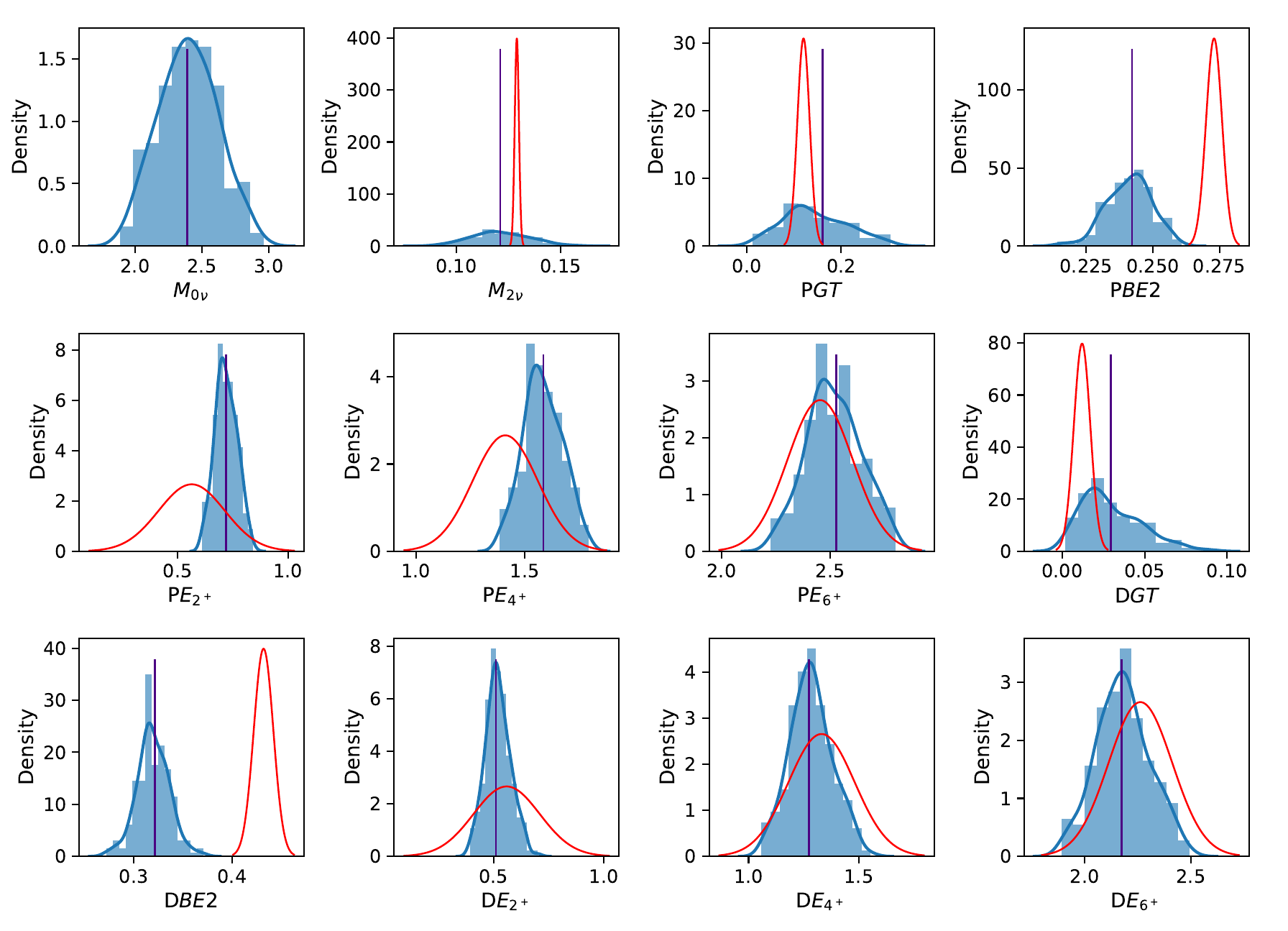}
\caption{Distributions based on experimental data (in red) compared with the KDE (in blue) obtained from the GCN2850 starting Hamiltonian. The green bars indicate the values of the observables for the starting effective Hamiltonians.}
\label{gcn-diag-dist}
\end{figure*}
\end{center}

\begin{center}
\begin{figure*}[htbp]
\includegraphics[ width=8.5cm]{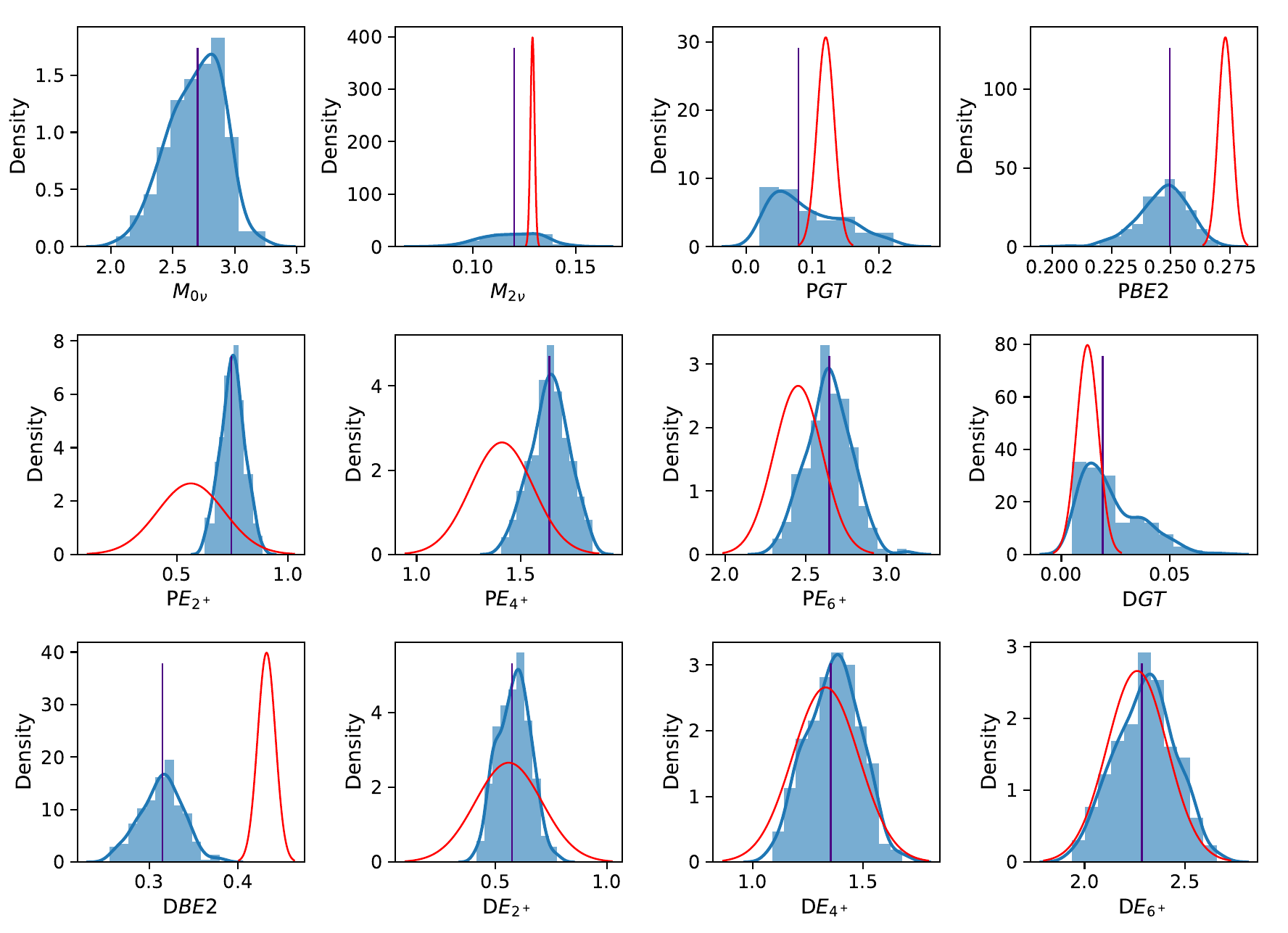}
\caption{Similar to Figure~\ref{gcn-diag-dist}, the distributions based on experimental data (in red) compared with the KDE (in blue) obtained from the JUN45 starting Hamiltonian.}
\label{jun45-diag-dist}
\end{figure*}
\end{center}

\begin{center}
\begin{figure*}[htbp]
\includegraphics[ width=8.5cm]{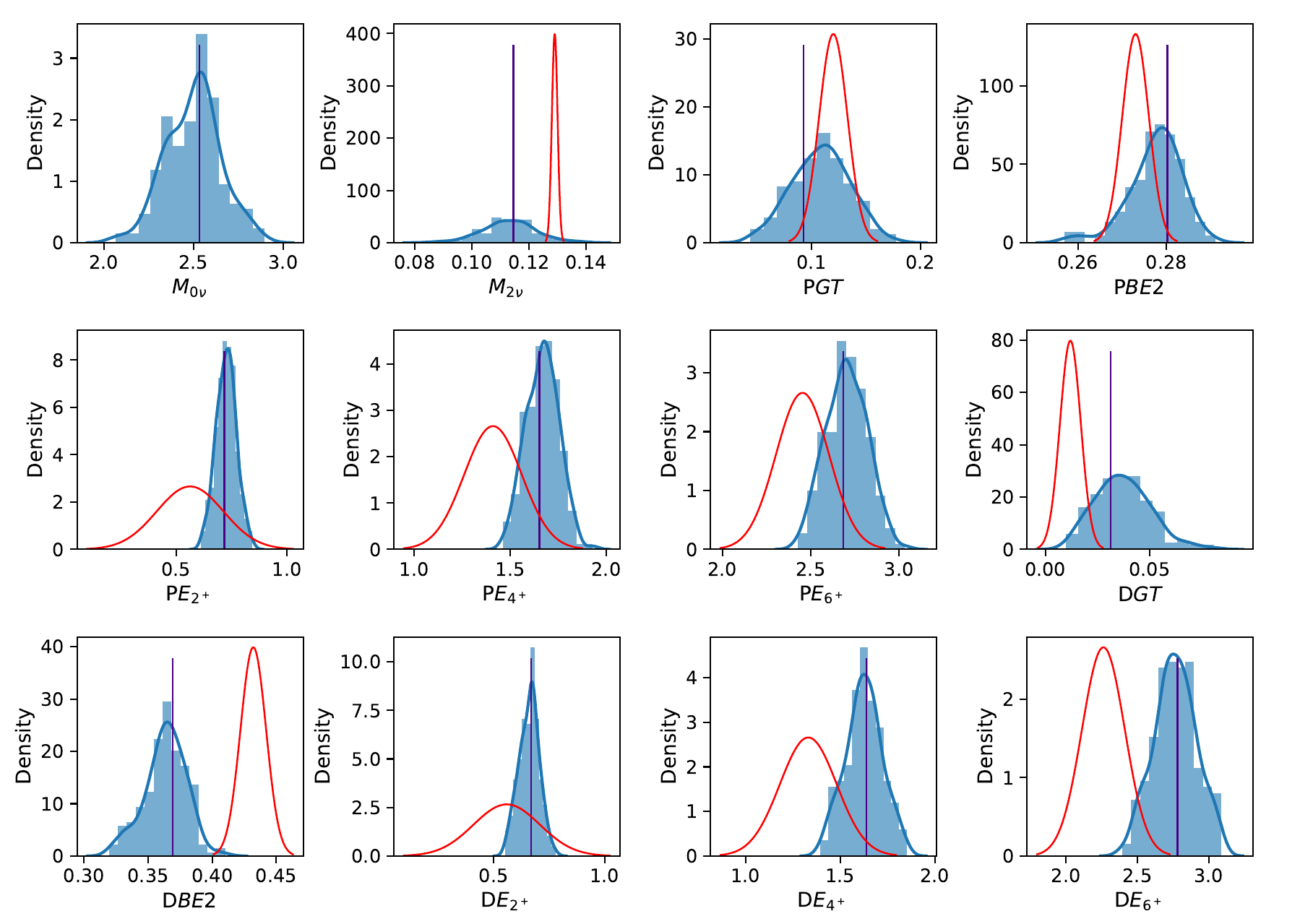}
\caption{Similar to Figures~\ref{gcn-diag-dist} and \ref{jun45-diag-dist}, the distributions based on experimental data (in red) compared with the KDE (in blue) obtained from the JJ44b starting Hamiltonian.}
\label{jj44-diag-dist}
\end{figure*}
\end{center}

Figure~\ref{gcn-diag-dist} overlays kernel density estimates (KDE, blue) of the ensemble distributions against experimental benchmarks (red), with vertical green markers indicating unperturbed Hamiltonian results. The red bands represent experimental uncertainties from Table~\ref{allstats}. Results for JUN45 and JJ44b (Figures~\ref{jun45-diag-dist} and \ref{jj44-diag-dist}) display nearly identical statistical behavior, confirming that the choice of baseline interaction does not significantly alter the ensemble topology. Minor deviations appear in GT probabilities, where TBME fluctuations can drive values toward zero; however, experimental data also occupy this low-probability region, leaving minimal margin for theoretical failure. As with other observables, B(E2)$\uparrow$ calculations using canonical effective charges do not fully overlap with experimental values, a known limitation of the $jj44$ space.

The authors of Ref.~\cite{JUN45} highlight that nuclei in the $jj44$ space frequently exhibit deformation, leading to enhanced B(E2)$\uparrow$ values along low-lying bands. The absence of $f_{7/2}$ and $g_{7/2}$ orbitals limits the model's capacity to fully capture quadrupole collectivity~\cite{Zuker1995}. While adjusting effective charges to $e_p=1.5$ and $e_n=1.1$ can improve B(E2) agreement, such modifications fall outside the scope of this uncertainty quantification effort. They do not alter correlation structures or NME conclusions, and we verify that the canonical charges yield consistent behavior across all three Hamiltonians. For reference, modified B(E2)$\uparrow$ values are: $PBE2_{gcn_s}=0.2127$, $PBE2_{jun_s}=0.2075$, $PBE2_{jj4_s}=0.2291$, and $DBE2_{gcn_s}=0.2311$, $DBE2_{jun_s}=0.2446$, $DBE2_{jj4_s}=0.3099$.

Following the protocol of Ref.~\cite{Horoi-Xe-2023}, we implement Bayesian Model Averaging to synthesize the three Hamiltonian ensembles into a unified probability distribution for $M_{0\nu}$. The posterior PDF is constructed as:

\begin{widetext}
\begin{equation}
\label{superposition}
\begin{split}
P(M_{0\nu})   = & W_{GCN2850} P_{GCN2850}(M_{0\nu}) + W_{JUN45} P_{JUN45}(M_{0\nu}) 
  +  W_{JJ44b} P_{JJ44b}(M_{0\nu}) \ .
\end{split}
\end{equation}
\end{widetext}

Evidence integrals~\cite{Horoi-Xe-2023,Horoi-universe-2024} yield posterior weights of $(1, 0, 0)$ for (GCN2850, JUN45, JJ44b), primarily due to GCN2850's superior spectral reproduction and JJ44b's underperformance in GT transitions. To avoid overconfidence in a single interaction, we adopt a compromise weighting scheme that averages prior (1/3 each) and posterior probabilities, resulting in final weights of $W_{GCN2850}=4/6$ and $W_{JUN45}=W_{JJ44b}=1/6$. Since the $M_{0\nu}$ distributions are tightly clustered and GCN2850 lies centrally between the other two, the weighted sum closely mirrors the GCN2850 KDE. This outcome suggests that, within the $jj44$ space, Hamiltonian selection exerts less influence on NME dispersion than in other model spaces where missing spin-orbit partners can shift values by $\sim20\%$~\cite{SenkovHoroiBrown2014,NeacsuHoroi2015,NeacsuHoroi2016}, a fraction far smaller than the factor-of-5 spread observed for $^{130}$Te~\cite{EngelMenendez2017,Rodejohann2019review}.

Figure~\ref{nme-range} displays the resulting PDFs alongside individual ensemble KDEs. The weighted sum (red curve) is normalized to unity and constructed via linear combination of the three KDEs, which were generated using Python's Pandas library. The mean and standard deviation of the weighted distribution correspond to the expectation value and theoretical uncertainty of $M_{0\nu}$. A 90\% confidence interval is obtained by integrating the PDF from both tails until each contains 5\% probability. The dimensionless NME units are preserved, and the PDF normalization ensures proper probabilistic interpretation.

\begin{widetext}
\begin{center}
\begin{figure}[H]
\includegraphics[width=\linewidth]{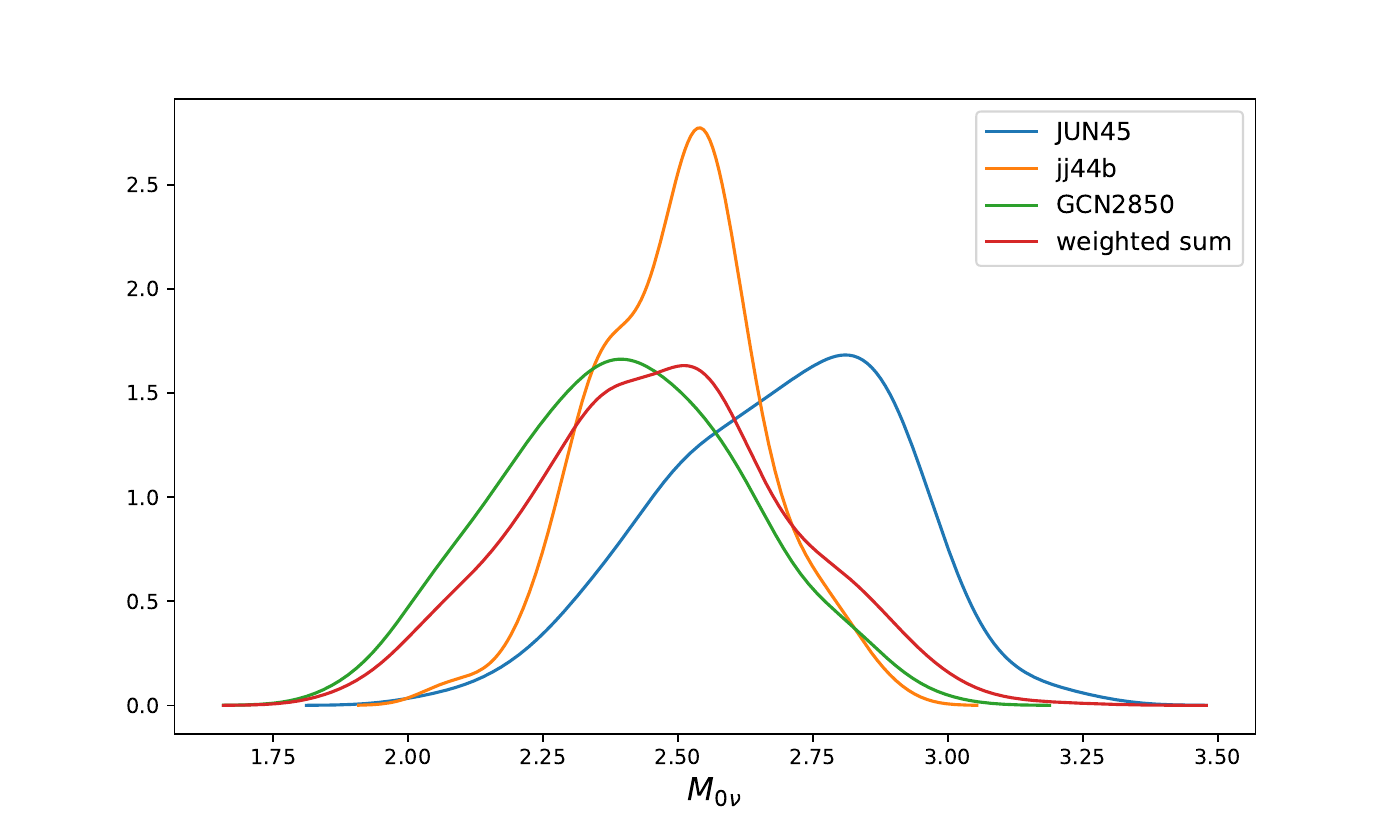}
\caption{PDFs of the $0\nu\beta\beta$ NME distributions for the JUN45, jj44b and GCN2850 Hamiltonians and their weighted sum (red curve, see text for details).}
\label{nme-range}
\end{figure}
\end{center}
\end{widetext}

The substantial overlap among all distributions confirms that shell-model calculations yield highly consistent $0\nu\beta\beta$ NME predictions in the $jj44$ space, irrespective of baseline interaction choice or minor optimization variations. This contrasts sharply with pn-QRPA results~\cite{EngelMenendez2017}, where parameter tuning (e.g., $g_{pp}$) often produces widely divergent outcomes. The weighted PDF can now serve as a rigorous theoretical prior for extracting neutrino mass limits from experimental half-life bounds, enabling more accurate isotope-mass scaling and resource allocation for next-generation detectors targeting the normal mass hierarchy. Additionally, this distribution can replace ad hoc error bars when comparing calculations that vary short-range correlation models, finite-size corrections, or higher-order current contributions.

\section{Conclusions and Outlook} \label{conclusions}

This work presents a comprehensive statistical quantification of the $0\nu\beta\beta$ NME uncertainty for $^{76}$Ge, leveraging the interacting shell model within the $jj44$ valence space ($f_{5}pg_{9}$). Three established effective Hamiltonians (GCN2850, JUN45, JJ44b) were perturbed by $\pm10\%$ in their TBME, generating ensembles of 200 interactions each. For every Hamiltonian variant, we computed twelve experimentally accessible observables, enabling a multi-dimensional consistency check and correlation analysis. The Bayesian Model Averaging framework synthesized these ensembles into a unified probability distribution, yielding a central NME value of 2.46 with a standard deviation of 0.25, corresponding to a 90\% confidence interval of $[1.89, 3.07]$.

Key findings include: (i) a robust positive correlation ($\rho > 0.9$) between $M_{0\nu}$ and $M_{2\nu}$, consistent across all isotopes studied in this statistical series; (ii) strong interdependencies among excited-state energies and between NMEs and quadrupole transition strengths; (iii) notable anti-correlations involving parent B(E2)$\uparrow$ values, reflecting underlying structural trade-offs; and (iv) remarkable stability of shell-model predictions across different baseline interactions, contrasting with the parameter sensitivity typical of QRPA approaches.

The mild dependence on the GT quenching factor ($q=0.65$) used for $2\nu\beta\beta$ and Gamow-Teller calculations further supports the reliability of our $0\nu\beta\beta$ results, which were computed without quenching under the closure approximation. Future ab-initio and many-body methods~\cite{Yao2020,Novario2021,Belley2021} may eventually eliminate the need for phenomenological quenching by deriving consistent effective operators from first principles. Our statistical framework provides a valuable benchmark for such efforts, enabling systematic uncertainty propagation and correlation mapping.

Beyond $0\nu\beta\beta$ decay, this methodology can be extended to other lepton-number-conserving and violating processes of experimental interest, such as ordinary muon capture~\cite{Kortelainen_2002,KORTELAINEN2003501,Zinatulina2019}, which is actively pursued by MONUMENT~\cite{bajpai2024monument}, LEGEND~\cite{legend2021}, and nEXO~\cite{nEXO_2022}. By quantifying theoretical uncertainties with controlled statistical ensembles, we aim to bridge the gap between nuclear structure theory and precision neutrino physics, ultimately guiding the design of next-generation experiments capable of probing the normal mass hierarchy and fundamental symmetry violations.

\vspace{0.3cm}
\begin{acknowledgments}
This research was funded by the Romanian Ministry of Research, Innovation and Digitization grant PNRR-I8/C9-CF264, Contract No. 760100/ 23.05.2023. M.H. acknowledges support from the US Department of Energy grant DE-SC0022538 ``Nuclear Astrophysics and Fundamental Symmetries''.
The authors acknowledge valuable technical support, resources, and runtime on the CIFRA \& National Institute of Materials Physics joint HPC cluster in Magurele, Romania.
\end{acknowledgments}

\bibliographystyle{apsrev4-2} 
\bibliography{ge76}

\end{document}